\documentstyle[aps,prl,psfig]{revtex}  

\begin{document} 
\draft 
\newcommand{\no}{\nonumber} 
\newcommand{\beq}{\begin{equation}} 
\newcommand{\eeq}{\end{equation}} 
\newcommand{\beqa}{\begin{eqnarray}} 
\newcommand{\eeqa}{\end{eqnarray}}
 
\title{Conductances in normal and normal-superconductor structures}

\author{F. Sols and J. S\'anchez-Ca\~nizares} 
\bigskip 
\address{ 
Departamento de F\'{\i}sica Te\'orica de la Materia Condensada, C-V, 
and\\ 
Instituto Universitario de Ciencia de Materiales ``Nicol\'as 
Cabrera''\\ 
Universidad Aut\'onoma de Madrid, E-28049 Madrid, Spain}

\maketitle 
\begin{abstract}

We study theoretically electronic transport through a normal metal -- 
superconductor (NS) interface and show that more than
one conductance may be defined, depending on the pair of chemical potentials 
whose difference one chooses 
to relate linearly to the current. We argue that the situation is analogous
to 
that found for purely normal transport, where different conductance
formulae can 
be invoked. We revisit the problem of the ``right" conductance formula in a 
simple language, and analyze its extension to the case of mesoscopic 
superconductivity. The well-known result that the standard conductance of a
NS 
interface becomes 2 (in units of $2e^2/h$) in the transmissive limit, is
viewed 
here in a different light. We show that it is not directly related to the 
presence of Andreev reflection, but rather to a particular choice of chemical 
potentials. This value of 2 is measurable because only one single-contact 
resistance is involved in a typical experimental setup, in contrast with the 
purely normal case, where two of them intervene. We introduce an 
alternative NS conductance that diverges in the transmissive limit due to the 
inability of Andreev reflection to generate a voltage drop. We illustrate 
numerically how different choices of chemical potential can yield widely 
differing I--V curves for a given NS interface.


\end{abstract} 

\vspace{.5cm} 

 
\section{Introduction}

Since the early work by Landauer \cite{land57}, the scattering picture has
provided a useful framework for theoretical studies on electron
transport in small structures. In the eighties, a debated question
\cite{ston88,land89} was whether the (zero temperature) conductance formula
for 
a barrier in a one-dimensional wire is
\beq
\label{gt}
G=TG_0
\eeq
or, rather,
\beq
\label{gtr}
\bar{G}=\frac{T}{R}G_0,
\eeq
where $G_0\equiv 2e^2/h$ and $T=1-R$ is the probability that an electron is 
transmitted across the barrier, as indicated in Fig. 1a.
The consensus
emerged that Eq. (\ref{gt}) is relevant as a two-lead conductance formula,
while Eq. (\ref{gtr}) should better describe a four-lead conductance. Eqs. 
(\ref{gt}) and (\ref{gtr}) were generalized to the
multi-channel case by B\"uttiker {\it et al.} \cite{buet85}, and a multi-lead 
extension of Eq. (\ref{gt}) was
derived by B\"uttiker \cite{buet86,buet88}.

In this article, we wish to address the problem of the ``right''
conductance formula in the context of mesoscopic superconductivity. We will
argue that, what is usually presented as the definition of conductance
in a normal-superconductor (NS) interface, is just a particular (albeit
rather 
natural) choice. We begin by reviewing the work of Ref. \cite{buet85} in a 
slightly different
language which will permit a convenient generalization to the NS case. For 
greater clarity, we focus on one-dimensional transport, neglecting the 
complication that, in reality, a stable superconductor requires higher 
dimensions.

We adopt the point of view that the two different conductance formulae
(\ref{gt}) and (\ref{gtr}) are equally valid for a given sample, the
difference lying in the choice of chemical potentials to which the electric
current is linearly  related. Normal transport requires the existence of 
differences among the chemical potentials of several subsets of carriers.
Hence it should come as no surprise that, if the same electric
current can be related to more than one pair of chemical potentials, it is
possible to formulate more than one definition of conductance. If one
reviews a standard derivation of Eq. (\ref{gt}) (see for instance Refs.
\cite{buet86,buet88}), one may note that the chemical potentials which are invoked are those which characterize the population of electrons in the {\it
incoming} channels. In terms of these chemical potentials, the electric
current 
$I$ can be calculated to be
\beq
\label{in-in}
\frac{hI}{2e}=T(\mu^{\rm in}_{\rm L} - \mu^{\rm in}_{\rm R}),
\eeq
where $\mu^{\rm in}_{\rm L}$ ($\mu^{\rm in}_{\rm R}$) is the chemical
potential 
for
electrons impinging on the sample from the left (right) lead, as
schematically 
shown in Fig. 1a. One readily notes that the conductance formula (\ref{gt}) 
results from defining \cite{comm10}
\beq
\label{g-in-in}
G\equiv \frac{eI}{\mu_{\rm L}^{\rm in}-\mu_{\rm R}^{\rm in}}.
\eeq
A standard assumption
is that both leads are connected to broad
reservoirs through ideal contacts lacking internal reflection, as depicted in 
Fig. 2a.
This permits us to assert that all incident electrons come directly from the 
reservoirs, those in each lead possessing an internal thermal distribution
which 
is identical to that of their original reservoir (which, by construction,
can be 
assumed to be in equilibrium). In particular, we can write $\mu_{\rm L}^{\rm 
in}=\mu_1$ and
$\mu_{\rm R}^{\rm in}=\mu_2$, where $\mu_1$ and $\mu_2$ are the chemical 
potentials of the left and right reservoirs, respectively.
This identification of the chemical potentials of the incoming electrons with 
those of the broad reservoirs where voltage is eventually measured, justifies 
the label ``two-lead conductance" for Eq. (\ref{gt}).

Unlike the incident electrons, one may expect the outgoing electrons in a
given 
lead to incorporate properties from both reservoirs. Assuming that the 
population of electrons emerging from the sample can be described
appropriately 
by chemical potentials $\mu_{\rm L}^{\rm out}$ and $\mu_{\rm R}^{\rm out}$
(see 
Fig. 1a), one expects the following relations to hold:
\beqa
\label{ansatz}
\mu_{\rm L}^{\rm out} &=& R \mu_{\rm L}^{\rm in} + T \mu_{\rm R}^{\rm in} 
\nonumber \\
\mu_{\rm R}^{\rm out} &=& T \mu_{\rm L}^{\rm in} + R \mu_{\rm R}^{\rm in}.
\eeqa
Eq. (\ref{ansatz}) provides an intuitive ansatz for the outgoing chemical
potentials in terms of the incoming chemical potentials which explicitly
invokes the scattering properties of the sample. In Appendix A, we provide
a rigorous derivation of Eq. (\ref{ansatz}) and explicitly prove its
equivalence to related equations derived in Ref. \cite{buet85}.

One may define an {\it average} chemical potential for each lead,
\beq
\label{average}
\bar{\mu}_{\alpha}= (\mu_{\alpha}^{\rm in}+\mu_{\alpha}^{\rm out})/2,
\;\;\;\; \alpha={\rm L,R},
\eeq
which is directly associated to the total electron density. It is easy to
prove 
that
\beq
\label{connection}
\bar{\mu}_{\rm L}-\bar{\mu}_{\rm R} =R(\mu^{\rm in}_{\rm L} -\mu^{\rm
in}_{\rm 
R}),
\eeq
which permits to rewrite (\ref{in-in}) as
\beq
\label{av-av}
\frac{hI}{2e}= \frac{T}{R}(\bar{\mu}_{\rm L} - \bar{\mu}_{\rm R}) .
\eeq
This result indicates that $\bar{G}$ in Eq. (\ref{gtr}) is
obtained from relating the electric current to differences between the
{\it average} chemical potentials associated to the total electron density in 
each lead, i.e., we can define
\beq
\label{g-av-av}
\bar{G}\equiv \frac{eI}{\bar{\mu}_{\rm L} - \bar{\mu}_{\rm R}}
\eeq
and obtain (\ref{gtr}). Since $\bar{\mu}_{\alpha}/e$ is the voltage
one would measure by attaching noninvasive capacitive (or, in general, weakly 
coupled) probes to both sides
of the sample \cite{land89,buet88,buet89,comm51}, hence the label ``four-probe conductance" some
times employed for Eq. (\ref{gtr}).

Within this simple approach to the problem, the choice of conductance formula 
reduces to the selection of two particular subsets of electrons whose
chemical 
potential
difference is linearly related to the current. There is no strict need to 
explicitly invoke 
concepts such as reservoirs or capacitive probes. The whole conductance
problem 
can be
formulated in terms of intrinsic scattering concepts. A different question is
what is the conductance that one actually measures in a particular
experimental 
setup \cite{land89} or that is relevant in a specific theoretical context.

\section{Conductances in a normal-superconductor interface}

Now we wish to extend this discussion of the conductance problem to the
case of a normal metal -- superconductor interface. A scattering approach to
electron transport through NS interfaces was already advocated by
Demers and Griffin \cite{deme70}, and by Blonder {\it et al.} \cite{blon82}, 
long  before such a viewpoint became popular in theoretical
studies of normal transport. The systematic extension of ideas on normal
mesoscopic transport to systems involving both normal and
superconducting elements was initiated by Lambert \cite{lamb91}, Beenakker
\cite{been92}, and Takane and Ebisawa \cite{taka92,hekk95}. However,
neither in 
these nor in other ensuing works has an explicit discussion been presented
of the conductance problem in NS interfaces \cite{comm53,lamb93,lamb98}. Rather, the view has
implicitly been adopted that there is only one NS conductance, which is the
natural generalization of Eq. (\ref{gt}). 
For the sake of simplicity, we focus on the
low-voltage regime in which normal and Andreev reflection are the only
outgoing scattering channels for quasiparticles impinging from the normal
lead N on the superconductor S, in the structures shown schematically in
Figs. 
1b and 2c. Quasiparticle number
conservation requires $A+B=1$, where $A$ ($B$) is the probability for Andreev
(normal) reflection from the NS interface.
A microscopic calculation shows that
\beq
\label{in-s}
\frac{hI}{2e}= 2A(\mu^{\rm in}_{\rm N}-\mu_{\rm S}) ,
\eeq
where $\mu_{\rm N}^{\rm in}$ is the chemical potential for incoming 
quasiparticles on the N side. At sufficiently low temperatures and voltages, 
there are no quasiparticles in S, which is thus described by a single
chemical 
potential $\mu_{\rm S}$. Eq. (\ref{in-s}) yields the famous NS conductance 
formula \cite{blon82,lamb91,been92,taka92}
\beq
\label{gnsa}
G_{\rm NS}\equiv \frac{eI}{\mu^{\rm in}_{\rm N}-\mu_{\rm S}}=2A G_0.
\eeq
A wave
function matching calculation shows that, at low energies
\cite{blon82,been92},
\beq
\label{att}
A=\frac{T^2}{(2-T)^2},
\eeq
where $T$ is the probability for electron transmission in the normal state of 
the interface. Eq. (\ref{att}) tells us that, in the tranmissive ($T
\rightarrow 
1$) limit, Andreev reflection becomes the dominant process ($A \rightarrow
1$). 
This limit will be discussed in depth later.

Below we show that a second NS conductance other than (\ref{gnsa}) can be 
introduced which is reminiscent of Eqs. (\ref{gtr}) and (\ref{g-av-av}) for
the 
normal case. We begin by presenting an ansatz for the outgoing chemical 
potential that adapts (\ref{ansatz}) to a NS context.
Since there are no quasiparticles on the S side, we only have to deal with one
incoming and one outgoing channel, both on the N side (here we view electrons 
and holes indistinctly, as terms which describe the occupation of a unique
set 
of electron states). In Appendix A, we show that the single relation that 
expresses $\mu_{\rm N}^{\rm out}$ in terms of $\mu_{\rm N}^{\rm in}$ must
read 
(see Fig. 1b)
\beq
\label{ansatz2}
\mu^{\rm out}_{\rm N}-\mu_{\rm S}= (B-A)(\mu_{\rm N}^{\rm in}-\mu_{\rm S}).
\eeq
The presence of quasiparticles at nonzero temperatures or high enough
voltages 
would of course complicate the picture slightly. The negative sign
accompanying 
the Andreev reflection probability in the l.h.s. of Eq. (\ref{ansatz2})
reflects
the fact that, upon Andreev reflection, an incident electron is
converted into a hole which, having opposite charge, lowers the
chemical of the outgoing carriers in N by an amount that, in the average, is 
proportional to the difference $\mu_{\rm N}^{\rm in}-\mu_{\rm S}$. 

>From (\ref{average}) (with $\alpha={\rm N}$) and (\ref{ansatz2}), it follows 
that \cite{comm54}
\beq
\label{connection2}
\bar{\mu}_{\rm N} -\mu_{\rm S} = B (\mu_{\rm N}^{\rm in} - \mu_{\rm S}),
\eeq
and one may rewrite (\ref{in-s}) as
\beq
\label{av-0}
\frac{hI}{2e}= \frac{2A}{1-A}(\bar{\mu}_{\rm N}-\mu_{\rm S}) .
\eeq
This suggests the introduction of an alternative NS conductance
\beq
\label{gnsar}
\bar{G}_{\rm NS}\equiv \frac{eI}{\bar{\mu}_{\rm N}-\mu_{\rm
S}}=\frac{2A}{1-A} 
G_0.
\eeq
This NS conductance has in common with the ``four-lead" normal conductance 
(\ref{gtr}) that it diverges in the transmissive limit ($T,A \rightarrow 1$). 
If we identify the electrostatic potential on the N side with $\bar{\mu}_{\rm 
N}/e$,
it follows that {\it there is no voltage drop in a current-carrying
transmissive NS interface}. Unlike normal reflection, Andreev reflection
{\it per se} does not generate an electron potential drop at the interface.
The 
physical
reason is clear: In Andreev reflection processes, incident
quasiparticles are reflected with opposite charge and thus do not
contribute to a net accumulation of charge on the N side, nor to an average
chemical potential imbalance between the N and S sides. Like in a
normal conductor, a voltage drop at the NS interface can only be generated by
normal reflection processes.

In the $T=1$ limit, $G$ and $G_{\rm NS}$
become 1 and 2, respectively, in units of $G_0$. The value of 2 for $G_{\rm
NS}$ 
is some times attributed to the dominant presence of Andreev reflection
($A=1$); 
an incident electron which converts into a hole moving in the opposite
direction 
is said to make a double contribution to the current.
Here we propose a different point of view, according to which $G_{\rm
NS}/G_0=2$ is not directly related to Andreev reflection but rather to the
properties of the {\it contacts} linking the leads to the reservoirs. To prove
this assertion, let us return momentarily to the normal case and 
consider a third definition of normal
conductance which one might introduce by writing the current in terms of a 
combination of incoming and average $\mu$'s, namely,
\beq
\label{in-av}
\frac{hI}{2e}= \frac{2T}{2-T}(\mu^{\rm in}_{\rm L} - \bar{\mu}_{\rm R}) ,
\eeq
as can be derived from (\ref{in-in}), (\ref{ansatz}), and (\ref{average}).
Eq. (\ref{in-av}) suggests the introduction of the conductance
\beq
\label{gtt}
\tilde{G}\equiv \frac{eI}{\mu^{\rm in}_{\rm L} - \bar{\mu}_{\rm
R}}=\frac{2T}{2-
T}G_0,
\eeq
which has the interesting property that, when $T=1$, it also satisfies 
$\tilde{G}/G_0=2$,
and yet there is no trace of Andreev reflection in the system. This
choice of normal conductance, which may look somewhat artificial for a purely 
normal system, can instead
be quite natural -and in fact is used- for a NS interface. 
We have said that $G_{\rm NS}$ in (\ref{gnsa}) generalizes Eq. (\ref{gt})
to the 
NS case. However, it is even more precise to view $G_{\rm NS}$ as an
extension 
of $\tilde{G}$, since it relates the current to the difference $\mu_{\rm
N}^{\rm 
in}-\mu_{\rm S}$. Being the only chemical potential in the superconductor, 
$\mu_{\rm S}$ can be regarded in particular as the ``average" chemical 
potential, hence the strong analogy between (\ref{gnsa}) and (\ref{gtt}).
>From this analysis of the transmissive limit, we conclude that, locally,
and in 
what refers to time-averaged currents and voltages, {\it there is no
distinction 
between an imaginary dividing line in a perfect normal lead and a transparent 
normal-superconductor interface}. In such a limit, the analogous conductances 
$\bar{G}$ and $\bar{G}_{\rm NS}$ diverge, while $\tilde{G}$ and $G_{\rm NS}$ 
acquire a value of 2. 

The normal conductance $\tilde{G}$ defined in (\ref{gtt}) does not seem to be 
relevant in typical situations. 
It is (\ref{gt}) what is usually measured \cite{ston88}, and (\ref{gtr}) only 
under especial conditions \cite{land89}. We have established the approximate 
physical equivalence between (\ref{gtt}) and the standard NS conductance 
(\ref{gnsa}) in the transmissive limit. Must we conclude that (\ref{gnsa}) is 
also physically meaningless? The answer is no, and the reason lies in the 
different behavior at the contacts displayed by the structures to which 
(\ref{gnsa}) and (\ref{gtt}) typically apply.

\section{The role of the contacts}

Before discussing the effect of the contacts, it is important to note that
two 
resistances in series may be summed only when they both are referred to the 
average chemical potential in the intermediate
region between them. In particular, resistances given by the inverse of 
(\ref{gtr}) and (\ref{g-av-av}) can always be summed, while those obtained
from 
the inversion of
(\ref{gt}) and (\ref{g-in-in}) cannot. The additivity of resistances of the
type 
$\bar{G}^{-1}$ is fully consistent with the well-known property that the
ratio $R/T$ for a double barrier is additive \cite{datt95} if multiple scattering by the
two 
barriers is assumed to be incoherent \cite{comment1,gram97}.

By the same rule, resistances of the type given by the inverse of
(\ref{gnsa}) 
or (\ref{gtt}) can be summed once, yielding a resistance of the type
(\ref{gt}) 
and (\ref{g-in-in}). 
In particular, in the transmissive limit, when their value is 1/2 (see
section 
II), the sum of two of them yields in both cases the value of 1 which one 
expects for the resistance of a perfect normal lead or a transparent NSN structure
\cite{sanc97,hui93}, such as those shown in Figs. 2a (without barrier) and 2b (with transmissive interfaces).

In our language, it is easy to see that an ideal normal contact connecting a 
broad reservoir with  a one-dimensional lead contributes 1/2 to the total 
resistance \cite{imry86} (that which relates the chemical potentials in the 
reservoirs). 
It suffices to remember that $\mu_1$ in the reservoir can be identified with 
$\mu_{\rm L}^{\rm in}$ in the lead, and to note that, by construction,
\beq
\label{in-out}
\frac{hI}{2e}= \mu_{\rm L}^{\rm in}- \mu_{\rm L}^{\rm out}
\eeq
in the narrow lead. From (\ref{ansatz}) and (\ref{average}), it follows that
\beq
\label{uno-av}
\frac{hI}{2e}= 2(\mu_1- \bar{\mu}_{\rm L}) ,
\eeq 
hence the single-contact resistance of 1/2. Summing the contributions from
the 
two contacts, one obtains the well-known value of 1 for the total contact 
resistance of a perfectly transmitting channel \cite{imry86}. This would 
correspond to the case depicted in Fig. 2a for a perfect normal lead, or in
Fig. 
2b for a lead containing a superconducting segment with transmissive
interfaces.

Fig. 2c shows schematically a possible setup to measure the NS resistance. In 
such a structure, the NS interface acts as the bottle neck controlling the 
current. The narrow superconducting lead S runs into a wide 
superconducting reservoir S' which is ultimately connected to a broad normal 
lead N'' through an ample contact. It is reasonable to neglect the potential 
drop at the extended S'N'' interface, where current density is vanishingly
small. 
In a resistance measurement relating $I$ with $\mu_1 -\mu_2$, the dominant 
contributions will come from the interface and the contacts. The main
difference 
with the purely normal case (Fig. 2a) is that, in a NS conductance
measurement, 
{\it there is no voltage drop at 
the narrow-wide superconducting contact} 
(SS' in Fig. 2c). A voltage imbalance, which occurs naturally at a normal 
contact (see above), is forbidden between the condensates of S and S'
because it 
would require a time-variation of the relative phase which is energetically 
forbidden due to the rigidity of the macroscopic wave function. We conclude
that 
{\it there is only one single-contact resistance} operating in the 
structure of Fig. 2c. In the limit of a transmissive NS interface, this
results 
in a physically measurable value of 2 for the conductance of such a structure 
\cite{comm52,buet93}.

One can extend the argument to a structure such as that depicted in Fig. 2d, 
where no contact resistance is present, and conclude that one must measure a 
null total resistance. Of course, this is what should be expected for a
totally 
superconducting structure such as e.g. a superconductor interrupted by a 
Josephson link. This well-known result is viewed here as one extreme case
of a 
general scenario in which a fully normal structure (Fig. 2a) is the opposite 
extreme, and the hybrid NS setup of Fig. 2c is a characteristic intermediate 
case \cite{comm2}. We emphasize again that it is not necessary to invoke
Andreev 
reflection explicitly in order to explain why a NS interface yields a
measurable 
conductance of 2.

\section{Numerical illustration} 

In the second part of this paper, we show with specific examples how
different 
voltages (obtained from different definitions of the chemical potential) can 
give rise to different I--V curves. We focus on a NS structure of the type 
depicted in
Fig. 2c. For simplicity, the one-dimensional model of Ref. \cite{blon82} is 
used and zero temperature is assumed. A barrier of the form $H\delta(x)$
($x$ is 
the longitudinal coordinate) is introduced at
the NS interface, so that the dimensionless parameter $Z \equiv
mH/\hbar^2k_F$ 
measures 
the barrier scattering strength. Given a total potential drop $V$ across the 
structure [here $V \equiv (\mu^{\rm in}_{\rm N}-\mu_{\rm S})/e$], the flowing 
electric
current may be calculated as \cite{blon82,comm9}:
\beq
I=\frac{2e}{h}\int_0^{eV}dE \, [1+A(E)-B(E)],
\label{current}
\eeq
where the energies are referred to $\mu_{\rm S} \equiv 0$. In Appendix A, we 
show that, at zero temperature, the average chemical potential on the normal 
lead is
\beq
\bar{\mu} _{\rm N}=\frac{1}{2}\int_0^{eV}dE \, [1-A(E)+B(E)],
\label{vn}
\eeq
When $eV$ exceeds the superconducting gap $\Delta$, quasiparticles in the 
superconductor cause a variation in the chemical potential which is given by
\beq
\bar{\mu} _{\rm Q}=\frac{1}{2}\int_{\Delta}^{eV}dE \, [C(E)-D(E)],
\label{vs}
\eeq
as is shown in Appendix A. In Eq. (\ref{vs}) $C(E)$ and $D(E)$ are the 
probabilities that an incident electron is transmitted as a quasielectron 
(normal transmission) and a quasihole (Andreev transmission), respectively.

In Fig. 3a we present the I--V characteristics of the system for different 
values of $Z$. Solid lines coincide with those of Ref. \cite{blon82}
and correspond to a definition of voltage given by $V = \mu^{\rm in}_{\rm
N}/e$,
[in the spirit of Eq. (\ref{gnsa}) with $\mu_{\rm S}=0$], i.e., the 
total voltage drop across the structure. Dashed lines result from plotting
the 
current as a function of the difference $\bar{\mu} _{\rm N}-\bar{\mu} _{\rm
Q}$. 
This is the I--V curve that would be measured in a hypothetical 4-probe
measurement. In Fig. 3b we plot the average chemical potentials in the normal 
(dashed-dotted line) and superconducting (solid line) leads as a function of 
the total voltage drop $\mu^{\rm in}_{\rm N}/e$.
For $Z=0$, one finds no average voltage drop: $\bar{\mu}_{\rm
N}=\bar{\mu}_{\rm 
Q}$ and the two curves coincide. We have already discussed (after introducing 
$\bar{G}_{\rm NS}$) the zero voltage limit of this general result. Here we 
confirm and generalize the local equivalence in transport between fully 
transmissive NN and NS interfaces.
As $Z$ increases, a difference between  $\bar{\mu} _{\rm L}$ and $\bar{\mu} 
_{\rm S}$ arises which approaches the total voltage drop $V$ when $Z\gg 1$.
In 
other words, the I--V curves obtained with different definition of voltages 
tend to become indistinguishable from each other as the tunnel barrier
limit is 
reached, i.e., when the barrier resistance is much bigger than that of the 
contacts. 
This trend can be clearly appreciated in Fig. 3a. 

It is also interesting to note the different evolution of the average
chemical 
potentials
due to the energy dependence of the scattering probabilities at the NS 
interface.
For any value of $Z$, there is no quasiparticle transmission into
the superconductor as long as $eV < \Delta$, and therefore $\bar{\mu} 
_{\rm Q}=0$, as shown in Fig. 3b.
At these low voltages, the slope $d\bar{\mu}_{\rm N}/d\mu^{\rm in}_{\rm N}$ 
increases with $Z$ due to normal reflection, as can be seen in Fig. 3b.
When $eV > \Delta$, normal transmission of quasiparticles becomes possible,
and 
the well-known phenomenon of quasiparticle charge imbalance arises in the 
superconducting lead \cite{clar72}. This charge imbalance eventually
relaxes as 
quasiparticles enter the wide reservoir. On the other hand,
the slope of $\bar{\mu} _{\rm N}$ decreases as $V$ exceeds 
$\Delta/e$ because the onset of quasiparticle transmission reduces normal 
reflection.

Finally, we wish to pay attention to an interesting feature occurring in the 
vicinity of $eV=\Delta$. It is known \cite{blon82} that, regardless 
of the value of $Z$, the Andreev reflection probability $A(E)$ reaches a
maximum 
value of 1 at $E=\Delta$ (if condensate flow can be neglected \cite{sanc95}).
This effect appears in the curves of Fig. 3b as a short plateau at 
$eV=\Delta$ which widens as $Z\rightarrow 0$. The flat slope is
characteristic 
of  a voltage window where Andreev processes dominate. In the limit $Z 
\rightarrow 0$, the plateau extends over the entire range $eV<\Delta$. This 
flattening of the I--V curve happens because Andreev reflection is
essentially a 
two-particle transmission process and therefore cannot generate a voltage
drop.

\section{Conclusions}
We have analyzed the variety of chemical potentials that can be defined in a 
transport context, both in purely normal and in hybrid normal-superconductor 
structures. In the former case, we have revisited the problem of the ``right" 
conductance formula in rather simple terms, associating each of the 
possible formulae to a particular choice of chemical potentials. For the NS 
interface, we have noted that it is also possible to derive more than one 
conductance formula. In particular, we have shown that the standard NS 
conductance, which becomes 2 (in units of $2e^2/h$) in the transmissive
limit, 
is 
just one particular choice, another one existing which diverges in the same 
limit. This divergence reflects the fact that Andreev reflection, which 
dominates at low applied voltages in the transparent limit, does not
contribute 
to a drop of the average voltage at the interface. We have made the case that 
the limiting value of 2 for the standard NS conductance of very smooth 
interfaces is 
not directly due to Andreev reflection. It is rather caused by the
existence, in 
a typical experimental setup, of only one contact where a voltage drop
occurs, 
namely, that which connects the narrow and wide normal leads. The other
contact, 
which connects the narrow and wide superconducting leads, cannot host a
voltage 
drop because of phase rigidity. This stays in contrast with the purely normal 
case in which a voltage drop occurs at both contacts, increasing the total 
resistance of the structure from 1/2 to 1. The opposite limit is well
known, and 
corresponds to that of a purely superconducting lead with a local narrowing: 
Since no voltage drop exists at any of the two narrow-wide contacts, the
total 
resistance is zero. Finally, we have illustrated with specific numerical 
examples how the I--V characteristics of a given NS interface can be very 
sensitive to the choice of chemical potentials against which the current is 
plotted.

\acknowledgments 

This work has been supported by the Direcci\'on General de Investigaci\'on 
Cient\'{\i}fica y T\'ecnica under Grant No. PB96-0080-C02, and by the TMR 
Programme of the EU. One of us (J.S.C.) acknowledges the support from
Ministerio 
de Educaci\'on y Ciencia through a FPI fellowship.

\appendix

\section{Densities and chemical potentials}

First we analyze the case of normal transport through a barrier in order to 
derive Eq. (\ref{ansatz}) rigorously and to prove that our approach reduces
to 
that of Ref. \cite{buet85}. We may write the density of incoming and outgoing 
electrons on the left side of the barrier as
\beqa
\label{n-in-out}
n^{\rm in}_{\rm L} &=& \int dE \, \, g_{\rm N}(E) f(E- \mu_{\rm L}^{\rm in}
) \\
n^{\rm out}_{\rm L} &=& \int dE \, \, g_{\rm N}(E) [R(E) f(E- \mu_{\rm
L}^{\rm 
in} ) 
+T(E) f(E- \mu_{\rm R}^{\rm in} )],
\eeqa
and analogously for the right side. $g_{\rm N}(E)$ is the density of states
for 
two spins and one direction, and $f(E) \equiv [\exp(\beta E)+1]^{-1}$ is the 
Fermi-Dirac function. If we refer everything to an equilibrium chemical 
potential $\mu_0$, we have $ \bar{\mu}_{\rm L} = \mu_0 +
\delta \bar{\mu}_{\rm L} $, with
\beq
\label{d-mu-av}
\delta \bar{\mu}_{\rm L} = \left( \frac{\partial n}{\partial \mu} \right)
^{-1}
\delta n_{\rm L}
\eeq
sufficiently small, $ n_{\rm L} = n^{\rm in}_{\rm L} + n^{\rm out}_{\rm L} 
$ the total electron density on the L side, and
\beq
\label{n-mu}
\frac{\partial n}{\partial \mu} = 2 \int dE \, g_{\rm N}(E) \left(   
-\frac{\partial f_0}{\partial E} \right).
\eeq
Here $f_0(E) \equiv f(E-\mu_0)$ is the equilibrium distribution and the
factor 
of 2 accounts for the existence of two directions. Analogously, we may define the 
variation 
in the outgoing chemical potential as
\beq
\label{d-mu-out}
\delta \mu_{\rm L}^{\rm out} = 2 \left( \frac{\partial n}{\partial \mu}
\right) 
^{-1} \delta n_{\rm L}^{\rm out}.
\eeq
As compared with (\ref{d-mu-av}), the factor of 2 appears because 
(\ref{d-mu-out}) deals with electrons moving only in the outgoing
direction. The result is 
that, up 
to linear order in the variations, we can  write
\beq
\label{out-in-in}
\mu_{\rm L}^{\rm out} = \frac {  \int dE \, g_{\rm N}(E) ( -\partial f_0 / 
\partial E )
[R(E) \mu_{\rm L}^{\rm in} + T(E) \mu_{\rm R}^{\rm in}]  } 
{\int dE \, g_{\rm N}(E) (-\partial f_0/\partial E) } \, ,
\eeq
which at low temperatures reduces to the ansatz (\ref{ansatz}) in the main
text. 
>From Eqs. (\ref{d-mu-av}) and (\ref{out-in-in}), one can show 
that Eq. (\ref{average}) for $\bar{\mu}_{\rm L}$ applies at all temperatures.

Some simple algebra leads to the result
\beq
\label{mu-lr-av-in}
\bar{\mu}_{\rm L} -  \bar{\mu}_{\rm R} = \frac { \int dE \, g_{\rm N}(E) ( -
\partial f_0 / \partial E ) R(E)}
{  \int dE \, g_{\rm N}(E) (-\partial f_0/\partial E) }
( \mu_{\rm L}^{\rm in} -\mu_{\rm R}^{\rm in} ),  
\eeq
which is equivalent to Eq. (2.7) of Ref. \cite{buet85}, and 
which
in the zero temperature limit yields the basic relation (\ref{connection}).

Below we present a similar analysis for the NS interface. We do not always 
restrict ourselves to the linear response limit, since the numerical
analysis of 
section 
IV includes the case in which the applied voltage is comparable or greater
than 
the superconducting gap. If we analyze the variation of $ \mu_{\rm N}^{\rm
out} 
$ starting from equilibrium as the voltage $V \equiv (\mu_{\rm N}^{\rm in}-
\mu_{\rm S})/e $ increases, we arrive at the following result (hereafter 
$\mu_{\rm S} \equiv 0$)
\beq
\label{d-mun-out}
\mu_{\rm N}^{\rm out}(V) = e \int_{0}^{V} dV' \, \frac
{\int dE \, g_{\rm N}(E) [-\partial f(E-eV')/\partial E] [B(E)-A(E)]}
{\int dE \, g_{\rm N}(E) [-\partial f(E-\bar{\mu}_{\rm N}(V')) / \partial
E] }.
\eeq
The presence of $\bar{\mu}_{\rm N}(V')$ in the integrand of the denominator 
accounts for the fact that the equivalent of Eqs. (\ref{n-mu}) and 
(\ref{d-mu-out})  must now be referred to the particular value of the average 
chemical 
potential at $V'$. The linear response limit is obtained by assuming that
$V$ is 
sufficiently small to make the replacement $\int_{0}^{V} dV'\rightarrow V$ 
evaluating the 
integrand at $V'=0$. The result is
\beq
\label{mu-n-out}
\mu_{\rm N}^{\rm out}= \mu_{\rm N}^{\rm in} \,
\frac {\int dE \, g_{\rm N}(E) [-\partial f_0/\partial E] [B(E)-A(E)]}
{\int dE \, g_{\rm N}(E) [- \partial f_0 / \partial E] },
\eeq
which is analogous to Eq. (\ref{out-in-in}).
On the other hand, the zero temperature limit of (\ref{d-mun-out}) is
\beq
\label{mu-out-v}
\mu_{\rm N}^{\rm out}(V) = e\int_{0}^{V} dV' \, [B(eV')-A(eV')],
\eeq
where we use the property that $ g_{\rm N}(E) $ varies very slowly in 
the energy scale of interest.
For both low temperatures and voltages, we reproduce the ansatz
(\ref{ansatz2}) 
of the text.
 
To obtain the average chemical potential, we replace $B-A$ by $(1+B-A)/2$
in Eq.
(\ref{mu-out-v}), with the factor of 2 again accounting for an increased
density 
of states (the two directions are involved). This reproduces Eq. (\ref{vn})
of 
the text. On the other hand, noting that $\mu_{\rm N}^{\rm in}$ 
is simply obtained by replacing $B-A$ by 1 in the r.h.s. of
(\ref{d-mun-out}), 
we prove that Eq. (\ref{average}) of the text (with $\alpha={\rm N}$) is
valid at 
all temperatures and voltages. 

The case of the quasiparticle chemical potential $\bar{\mu}_{\rm Q}$ in the 
superconductor is slightly more involved, since, for its calculation, in 
addition to the obvious 
replacement of $(B-A)$ by $(C-D)/2$, the normal density of states $g_{\rm
N}(E)$ 
in (\ref{d-mun-out}) has to be substituted by the superconducting density of 
states $g_{\rm S}(E)$, which cannot be taken as approximately constant in the 
relevant energy interval. Moreover, the transmitted quasielectrons and 
quasiholes have fractional charge. Fortunately, these two complications
cancel 
\cite{blon82} and the final result is not more involved than its normal 
counterpart. Specifically, we obtain
\beq
\label{mus-q}
\bar{\mu}_{\rm Q} (V) = \frac{e}{2} \int_{\Delta/e}^{V} dV' \, \frac
{\int dE \, g_{\rm N}(E) [-\partial f(E-eV')/\partial E] [C(E)-D(E)]}
{\int dE \, g_{\rm N}(E) [-\partial f(E-\bar{\mu}_{\rm Q}(V')) / \partial
E] }.
\eeq
Invoking arguments similar to those used for its normal analog 
(\ref{d-mun-out}), one can prove that, in the limit of zero temperature, 
Eq. (\ref{mus-q}) 
reduces to Eq. (\ref{vs}) of the text. Note that, unlike (\ref{d-mun-out}),
Eq. 
(\ref{mus-q}) does not have a well-defined linear response limit, since it
only 
applies for $eV > \Delta$.


\begin{figure}
\psfig{file=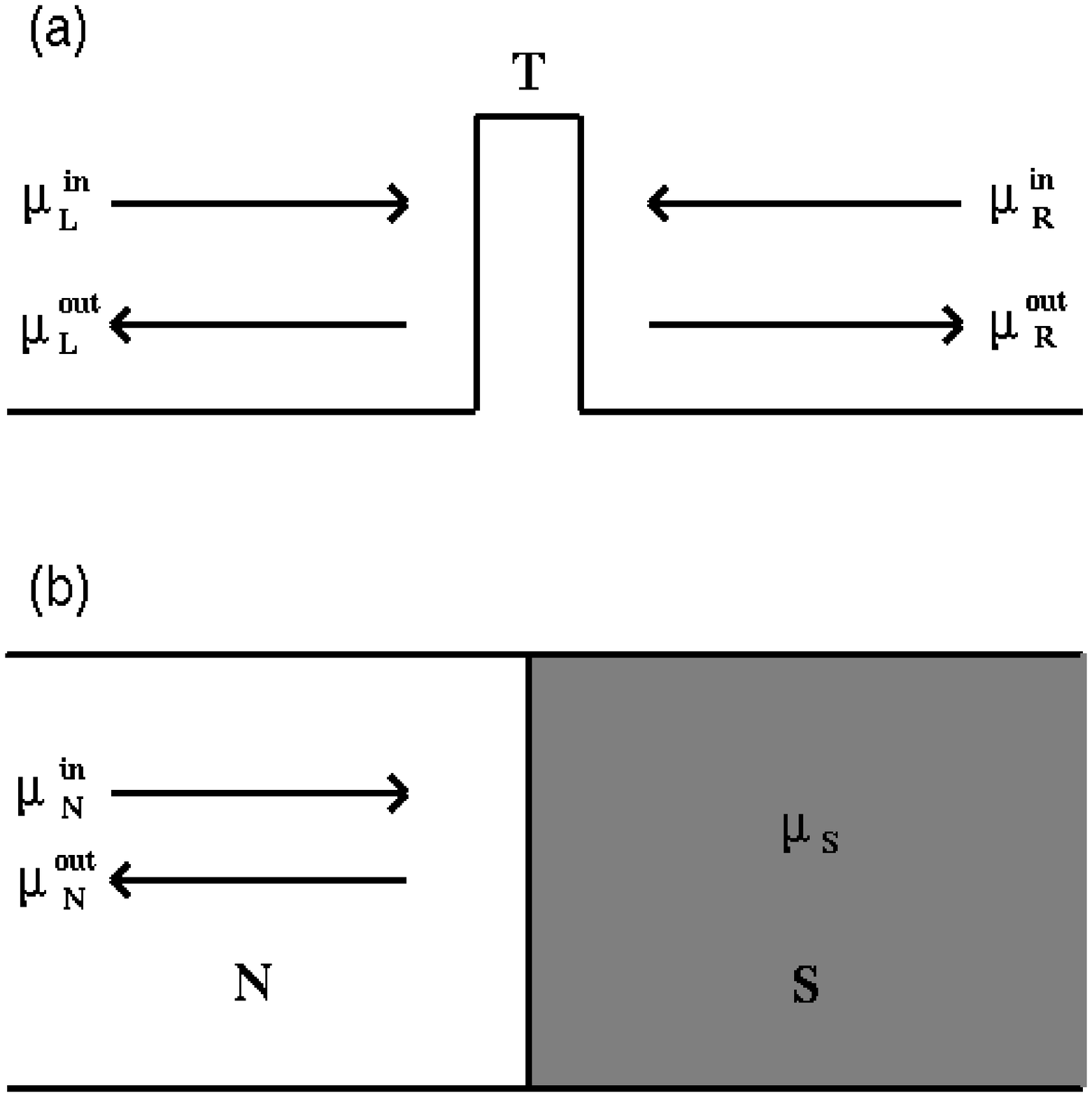,height=20cm,width=18cm,angle=0}
\caption{
(a) Normal transport through a barrier may be described in terms of
incoming and 
outgoing chemical potentials characterizing the populations of electrons
moving 
in the respective directions. $T$ is the probability for an electron to be 
transmitted across the barrier. 
(b) A similar picture may be adopted for an NS interface at low
temperatures and 
voltages, when the superconductor is characterized by a single chemical 
potential $\mu_{\rm S}$.
}
\end{figure}

\begin{figure}
\psfig{file=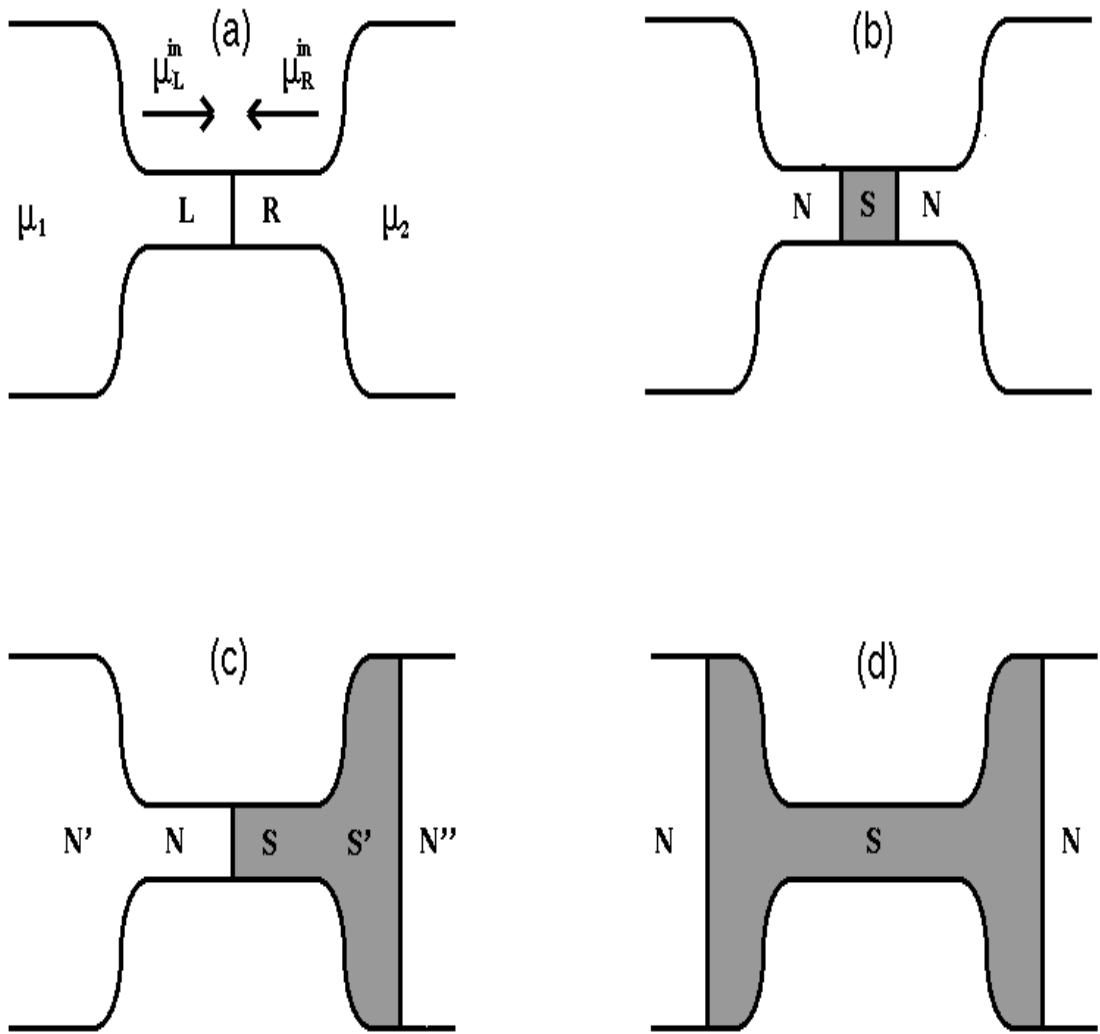,height=20cm,width=20cm,angle=270}
\caption{
(a) The lead containing the interface separating L and R is ultimately
connected 
through ideal contacts to wide reservoirs which are practically in
equilibrium 
and whose chemical potentials $\mu_1$ and $\mu_2$ can be identified, 
respectively, with 
the chemical potentials $\mu_{\rm L}^{\rm in}$ and $\mu_{\rm R}^{\rm in}$  
characterizing the incident electrons from the left and right leads. (b) A 
similar description holds if the barrier is replaced by a superconducting 
segment S with possibly reflecting NS interfaces. (c) In a typical setup to 
measure the NS resistance, the superconducting lead runs into a wide 
superconducting reservoir S' which is ultimately connected to a normal lead 
through an ample contact. (d) The case of a fully superconducting structure 
without net voltage drop is shown here to permit comparison with the cases 
depicted in (a-c).
}
\end{figure}

\begin{figure}
\psfig{file=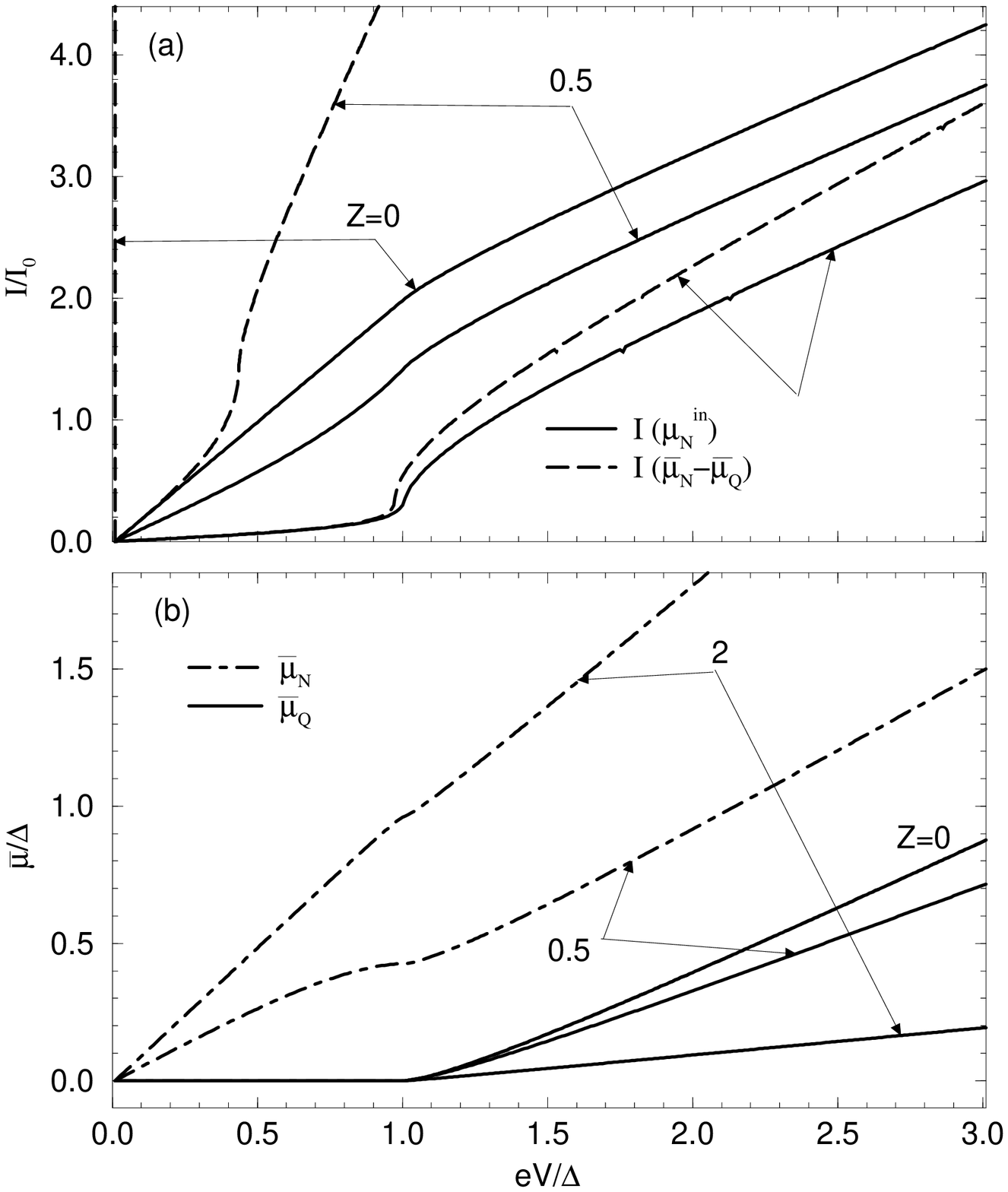,height=20cm,width=18cm,angle=0}
\caption{
(a) Electric current [in units of  $I_0 \equiv 2e \Delta/h(1+Z^2)$]
flowing through a NS interface plotted as a 
function of different chemical potentials for various values of the 
scattering strength $Z$. The solid lines reproduce the results of 
Ref.  8, since they are plotted against the chemical potential of the 
incident electrons, which is that of the reservoir. The dashed lines result
from 
plotting the currrent as a function of the average chemical 
potential drop $\bar{\mu}_{\rm N} - \bar{\mu}_{\rm Q}$.
(b) Average chemical potentials in the normal (dashed-dotted lines) and 
superconducting (solid lines) leads, plotted as a function of the total 
potential drop $V=\mu_{\rm N}^{\rm in}/e$.
}
\end{figure}
         

\begin{references}
\bibitem{comm10} Here we simply identify chemical potentials and voltages 
$\mu\equiv eV$, i.e., we adopt a noninteracting picture whereby $\mu$ and $V$ 
serve indistinctly to parametrise carrier populations.
\bibitem{comm51} This point can be proved rigorously. In particular, it can be shown that Eq. (1) of Ref. \cite{buet89} (giving the chemical potential at probe 3 in a three-port conductor when current flows between probes 1 and 2) reduces to our Eq. (\ref{average}) in the limit of weak and phase-incoherent coupling.
\bibitem{comm54} This result can be obtained as a particular case of Eq. (5) in (the first) Ref. \cite{lamb91} for a NSN structure, if one of the N's is replaced by a semi-infinite extension of S.
\bibitem{comment1} We are assuming that electron decoherence between the 
barriers takes place without any additional backscattering. This idealized 
situation is never strictly realized in
practice, but it provides a
useful coarse-grained description of mesoscopic resistances. It is
ultimately the combination of decoherence and multiple scattering which is
responsible for the ubiquity of Ohm's law in normal leads, according to which 
resistance
grows linearly with length. The difference between phase-coherent and phase-incoherent resistance measurements has been discussed in Refs. \cite{buet89} and \cite{gram97}.
\bibitem{comm52} There is here an interesting analogy with the behavior of charge relaxation resistances \cite{buet93}. When there is only one 
conductor-reservoir interface through which a conductor can relax its charge, then the effective conductance per channel is $2G_0$.
\bibitem{comm2} The effect of locally narrowing a superconducting lead
reduces 
to that of decreasing the critical current.
\bibitem{comm9} Like in Ref. \cite{blon82}, we neglect the effect of 
superconducting flow.
\end{references}
\end{document}